\documentclass[12pt,preprint]{aastex}
\usepackage{natbib}
\shorttitle{Measuring NIR Extinction with GPS}
\shortauthors{Blake and Shaw}

\begin{document}


\title{Measuring NIR Atmospheric Extinction Using a Global Positioning System Receiver}

\author{Cullen H. Blake and Margaret M. Shaw}
\affil{Princeton University Department of Astrophysical Sciences, Peyton Hall, Ivy Lane, Princeton, NJ 08544}

\begin{abstract}

Modeling molecular absorption by Earth's atmosphere is important for a wide range of astronomical observations, including broadband NIR photometry and high-resolution NIR spectroscopy. Using a line-by-line radiative transfer approach, we calculate theoretical transmission spectra in the deep red optical ($700<\lambda<1050$~nm) for Apache Point Observatory. In this region the spectrum is dominated by H$_{2}$O, which is known to be highly variable in concentration on short timescales. We fit our telluric models to high-resolution observations of A stars and estimate the relative optical depth of H$_{2}$O absorption under a wide range of observing conditions. We compare these optical depth estimates to simultaneous measurements of Precipitable Water Vapor (PWV) based on data from a Global Positioning System (GPS) receiver located at Apache Point. We find that measured PWV correlates strongly with the scaling of H$_2$O absorption lines in our spectra, indicating that GPS-based PWV measurements combined with atmospheric models may be a powerful tool for the real-time estimation of total molecular absorption in broad NIR bands. Using photometric measurements from the Sloan Digital Sky Survey (SDSS) DR8 database  
we demonstrate that PWV biases the calibrated r-z colors and z-band fluxes of mid-M stars but not mid-G stars. While this effect is small compared to other sources of noise in the SDSS z-band observations, future surveys like the Large Synoptic Survey Telescope aim for higher precision and will need to take time-variable molecular transmission into account for the global calibration of NIR measurements of objects having strong spectral features at these wavelengths. Empirical calibrations based on PWV may be immediately applicable to ongoing efforts to make mmag differential measurements of M stars to detect transiting exoplanets. 

\end{abstract}

\keywords{Astronomical Techniques}

\section{Introduction}

\begin{figure}
\plotone{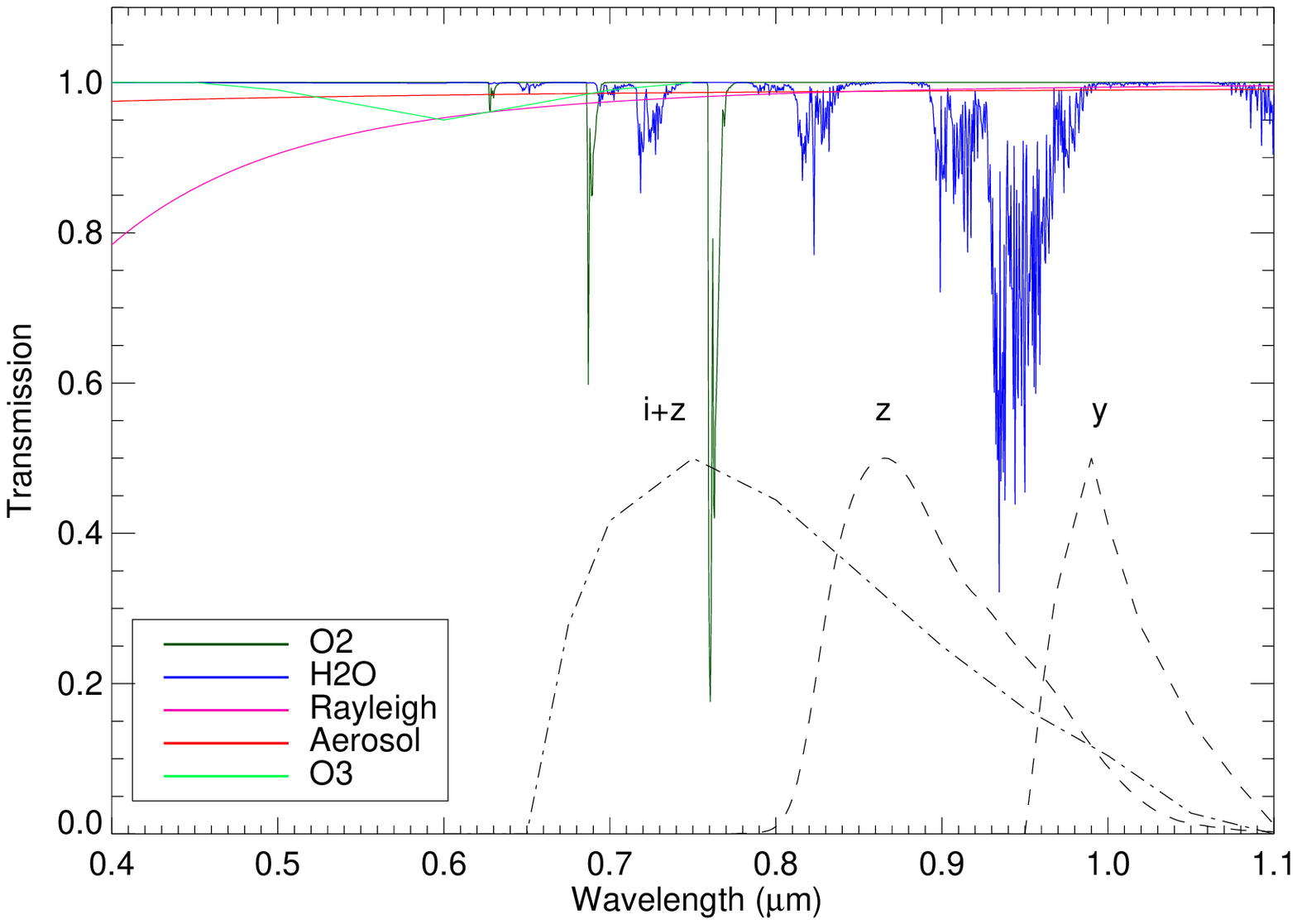}
\caption{Components of atmospheric absorption for a site like Apache Point Observatory (see also \citealt{burke2010}). The transmission curves for NIR bands are also shown, each normalized to a peak transmission of 50$\%$. The y band filter shown here is based on the proposed LSST y$_3$ filter \citep{high2010}.}\label{fig8}
\end{figure}

For many astronomical measurements it is necessary to understand and precisely measure the attenuation of light by 
Earth's atmosphere. While attempts have been made to do this in an absolute sense \citep{hayes1975}, in general astronomical measurements at optical and NIR wavelengths are fundamentally differential, measuring flux relative to either a standard star or comparison stars within a given field of view. As shown in Figure \ref{fig8}, across the optical and infrared Rayleigh scattering by molecules, Mie scattering by aerosols, and line absorption by molecules leads to atmospheric transmission that is a rapidly varying function of wavelength and can also vary rapidly in time \citep{stubbs2007}. Massive imaging surveys like the Large Synoptic Survey Telescope\footnote{\url http://www.lsst.org/}(LSST), Dark Energy Survey\footnote{\url http://www.darkenergysurvey.org/}(DES), Pan-STARRS\footnote{\url http://pan-starrs.ifa.hawaii.edu/public}, and Hyper Suprime-Cam\footnote{\url http://www.naoj.org/Projects/HSC/index.html} seek to produce broadband photometric measurements across the UV and near infrared (NIR) that are stable and uniform to better than $1\%$ over periods of years. Even if the response of the telescope to monochromatic light of a given intensity is known very precisely (e.g. \citealt{stubbs2006}, \citealt{stubbs2010}), global calibration of photometric measurements requires correcting flux for atmospheric variations. Similarly, precise radial velocity measurements at NIR wavelengths require a detailed understanding of atmospheric transmission so that subtle Doppler shifts in stellar spectral features can be separated from changes in telluric absorption features.

The flux in ADU measured by a single-element detector from an astronomical source, $\digamma$, can be expressed as a function of the wavelength-dependent properties of the telescope, atmosphere, and source as
\begin{equation}
\digamma = A \Delta T \int_{0}^{\infty} F_{\lambda} T(\lambda) S(\lambda) d\lambda
\end{equation}
where $A$ is the effective collecting area of the telescope, $\Delta T$ is the exposure time, $T(\lambda)$ is the transmission of the atmosphere as a function of wavelength, $S(\lambda)$ is the response function of the telescope in units of ADU/photon (including the effects of optics, filters, and detectors), and $F_{\lambda}$ is the flux density of the source in units of photons s$^{-1}$cm$^{-2}$cm$^{-1}$. Historically, for broadband photometry the band-integrated value of the product $S(\lambda)T(\lambda)$ is measured for a given night using a combination of observations of a uniformly illuminated screen (the ``flat field'') and observations of photometric ``standard'' stars. Using this methodology it is possible, under excellent observing conditions, to measure the absolute brightnesses of objects with a precision of $1\%$, though these measurements are only as accurate as the measurements of the standard stars themselves. 

Temporal changes in $T(\lambda)$ and $S(\lambda)$ become important when the target and standard are not observed simultaneously. For many applications, such as precise photometry of transiting extrasolar planet hosts, only the relative brightness of an object, as compared to other nearby stars measured simultaneously, is required. This \textit{differential} photometric approach results in such high photometric precision, often $0.1\%$ (e.g. \citealt{chan2011}), in part because the temporal variations in $T(\lambda)$ and $S(\lambda)$ are intrinsically removed. Given multiple observations of a sufficiently large number of objects, it is possible to generalize this differential approach and apply it to data from large-area surveys to internally calibrate the brightnesses (and colors) of objects. \citet{padmanabhan2008} applied a technique of this type called \textit{\"{U}bercal} to the Sloan Digital Sky Survey (SDSS) photometric measurements and demonstrated that $1\%$ global photometric precision can be achieved for a survey of millions of objects spanning years (see also \citealt{ivezic2007}). 

Differential calibration strategies may not account for the fact that $F_{\lambda}$ varies strongly with stellar spectral type (or quasar redshift), resulting in higher-order effects that can arise when measuring relative fluxes between objects having very different Spectral Energy Distributions (SEDs). This is particularly important at NIR wavelengths, where strong molecular bands due to H$_{2}$O, CH$_{4}$, and CO$_{2}$ dominate $T(\lambda)$ and $F_{\lambda}$ may have strong absorption or emission features \citep{bailer-jones2003, blake2008a, mann2011}. In these cases, temporal variations in $T(\lambda)$ are not effectively removed through differential measurement, and residuals at the level of 1 mmg are expected in differential photometry of cool stars. Ultimately, the highest precision will be achieved for a survey like LSST with the help of detailed models of $T(\lambda)$ that are contemporaneous with broadband photometric observations and localized to particular sight lines. \citet{burke2010} demonstrate an ambitious strategy whereby simultaneous spectrophotometry of calibration stars in LSST fields can be combined with radiative transfer calculations for the atmosphere above the observatory to measure all important components of $T(\lambda)$. 

From the observational point of view, the challenges of precise radial velocities at NIR wavelengths and the global calibration of NIR photometry in large surveys are closely related since both require contemporaneous modeling of transmission through Earth's atmosphere. Precise stellar radial velocity measurements at wavelengths longer than 600 nm require a detailed understanding of the impact of absorption due to molecules in Earth's atmosphere \citep{wang2011}. These telluric absorption features themselves can also be used as a simultaneous wavelength reference, a measurement technique conceptually similar to the $I_{2}$ cell described by \citet{butler1996}, though first proposed by \citet{griffin1973}. 
The use of telluric lines as a wavelength reference has a long history in the literature (e.g. \citealt{smith1982}, \citealt{deming1987}, \citealt{hatzes1993}), and recently several authors have demonstrated that telluric lines are a viable velocity reference for planet searches, including \citet{seifahrt2008}, \citet{blake2010}, \citet{figueira2010harps}, and \citet{figueira2010crires}. This technique requires an excellent model of $T(\lambda)$, which can be derived empirically from high-resolution observations of the Sun \citep{blake2010} or calculated directly based on assumptions about the composition and structure of Earth's atmosphere \citep{seifahrt2010}.

Atmospheric constituents like CH$_{4}$, CO$_{2}$, and O$_{2}$ are well-mixed from Earth's surface through the mesosphere and have seasonal variations in total column density. For wavelengths $900<\lambda<2500$~nm, absorption by H$_{2}$O becomes very important, having large optical depth at some wavelengths. Modeling these absorption features poses a significant challenge since the total water column can have large and rapid variations. We demonstrate that Precipitable Water Vapor (PWV) estimates derived from a Global Positioning System (GPS) receiver can be used to calculate theoretical transmission spectra having no free parameters and resulting in excellent fits to astronomical spectra in the wavelength range $900<\lambda<1000$~nm. In a region of non-saturated H$_2$O lines we find fit residuals $<2\%$, likely dominated by residual fringing signals in our spectra. We find that GPS-based PWV estimates are also useful for estimating total transmission in the SDSS z band and demonstrate that changes in PWV bias photometric measurements of mid-M stars and determine an empirical correction factor for the z-band photometric measurements in the SDSS database. 

\section{ARC Echelle Observations}

As part of an ongoing program to measure the radial velocities of M stars, we observed 74 A star telluric standards with the ARCES echelle spectrograph \citep{wang2003} on the 3.5~m telescope at Apache Point Observatory (APO). A single ARCES observation contains data in more than 100 spectral orders spanning 360~nm to 1010~nm at a resolution of R$\sim$30,000. These A star observations were collected during 2010 under a wide range of observing conditions and at airmass (AM) up to AM=2.0. Exposure times were tailored so that S/N per pixel in the extracted spectra would be $\sim500$ at $\lambda\sim800$~nm. The ARCES spectral orders are closely spaced and marginally sampled in the spatial direction, requiring special care in the spectral extraction process so as not to induce aliasing. We collected a large number of bias and quartz lamp frames on each night and used the quartz lamp frames to define the positions of the centers of each of 75 spectral orders, which run approximately parallel to the detector rows. After subtracting an average nightly bias, we estimated scattered light in both the stellar spectra and quartz lamp spectra by fitting a low-order polynomial surface to the 74 inter-order minima of each column. Next, we extracted the stellar and quartz lamp spectra using an aperture extraction algorithm with fractional pixel weighting at the edges of the aperture. Since the orders are so narrow, it was necessary to flat field the ARCES data using the 1D extracted quartz lamp spectra, so we made a master flat for each spectral order for each observing night by averaging the individually normalized 1D quartz spectra. For the red orders of interest here there is large amplitude fringing in the ARCES spectra, which is partially removed from the stellar spectra by dividing by the normalized master flats. There are slight ($\sim$1 pixel) offsets between the positions of the spectral orders in the quartz and stellar frames, resulting in incomplete fringe removal. 

\section{Atmospheric Models}

Modeling telluric H$_{2}$O absorption in our A star spectra requires high-resolution telluric templates. In \citet{blake2010} an observed atmospheric
transmission spectrum, derived from solar observations by \citet{livingston1991}, was scaled to fit telluric absorption features due to CH$_{4}$ around 2300~nm. We found this approach to be problematic for absorption by H$_{2}$O between 800 nm and 1000 nm, so instead we created theoretical transmission spectra. Closely following \citet{gordley1994} we developed a custom line-by-line radiative transfer code to calculate template spectra that contain only molecular absorption (no scattering, no line mixing, no H$_{2}$O continuum absorption) given an atmospheric model and molecular transition data from the HITRAN 2008 database \citep{rothman2009}. Following the Beer-Lambert law for extinction through a medium by a single transition of a single absorber, the transmission, $T_{\nu}$, of the Earth's atmosphere at frequency $\nu$ is given by
\begin{equation}
T_{\nu}=e^{-u\sigma(\nu)}=e^{-\tau_{\nu}}
\end{equation}
where $\sigma(\nu)$ is the frequency-dependent cross section and $u$ is the mass path
\begin{equation}
u=\frac{qPx}{kT}
\end{equation}
where $x$ is the path length through the atmosphere, $P$ is pressure, $T$ is temperature, $k$ is Boltzman's constant, and $q$ is the Volume Mixing Ration (VMR), which is defined as the fraction of the particles in a volume that are the given constituent. Earth's atmosphere contains many gasses and the properties of the atmosphere, including $P$ and $T$, change with altitude and time. The transmission through the atmosphere can be calculated by breaking the atmosphere up into a number of levels in the vertical direction, $z$, within which the bulk properties are assumed to be constant. The total transmission due to a number of different absorbing gasses, $g$, each with a number of absorption lines, $i$, can then be calculated
\begin{equation}
T_{\nu}=e^{-\sum_{z}\sum_{g}u_{g,z}\sum_{i}\sigma_{g,i,z}(\nu)}
\end{equation}

The shapes and strengths of the absorption lines depend on the cross section, $\sigma({\nu})$. Absorption lines in Earth's atmosphere have intrinsic shapes both due to the bulk properties of the gas mixture and the quantum mechanical properties of the transition. The approximate shape of an absorption line is given by the Voigt profile, $V(\nu)$, which is the convolution of line broadening kernels due to Doppler and pressure (Lorentzian) broadening mechanisms. At low pressure the motions of the absorbing molecules dominate the broadening and at higher pressure collisions between molecules come to dominate. Both regimes occur at different levels in Earth's atmosphere. The Voigt profile is defined
\begin{equation}
V(\nu)=\frac{1}{\alpha_{D}\pi^{3/2}}\frac{\alpha_{L}}{\alpha_{D}}\int \frac{e^{-t^2}}{\left[\left(\nu-\nu_{c}\right)/\alpha_{D}-t\right]^{2}+\left[\alpha_{L}/\alpha_{D}\right]^{2}} dt
\end{equation}
where $\alpha_{D}$ is the Doppler half width of the line, $\alpha_{L}$ is the Lorentzian half width of the line, and $\nu_{c}$ is the line center, each of which can depend on atmospheric temperature and/or pressure.  The cross section $\sigma(\nu)$ is a function of the line broadening as well as the line intensity, $S=S^{0}Q$, where $Q$ is a scaling factor based on the partition function of the molecule and the temperature of the gas. The total transmission can now be expressed as $T_{\nu}=e^{-\tau_{\nu}}$ where $\tau_{\nu}$ is 
\begin{equation}
\tau_{\nu}=\sum_{z}\sum_{g}u_{g,z}\sum_{i}V_{g,i,z}(\nu)S_{g,i,z}
\end{equation}
We retrieved line parameters including half-widths, strengths, line centers, and pressure shifts of those line centers, for transitions of H$_{2}$O in the wavelength range 600 to 11000 nm from HITRAN 2008 and calculated atmospheric transmission at zenith in wavelength steps of 0.01\AA~across a plane parallel atmosphere with 30 layers logarithmically spaced between 0 km and 80 km.

This radiative transfer calculation requires a model of the structure and composition of Earth's atmosphere, including temperature, pressure, and chemical composition, as a function of height. We used the NASA-MSFC Earth Global Reference Atmospheric Model (Earth-GRAM 2010; \citealt{justus2004} and references therein) to approximate the average properties of the atmosphere above APO, located in southeastern New Mexico at an altitude of 2,780~m. Earth-GRAM combines a large number of historic atmospheric measurements with modern meteorological models to predict atmospheric properties, and the statistical distributions of those properties, as a function of altitude, location on the Earth, and month. While Earth-GRAM is often used for engineering applications, such as estimating the effect of Earth's atmosphere on trajectories, it is also a powerful resource for quantifying the impact of the atmosphere on astronomical observations. 

\section{Spectral Fitting}

\begin{figure}
\plotone{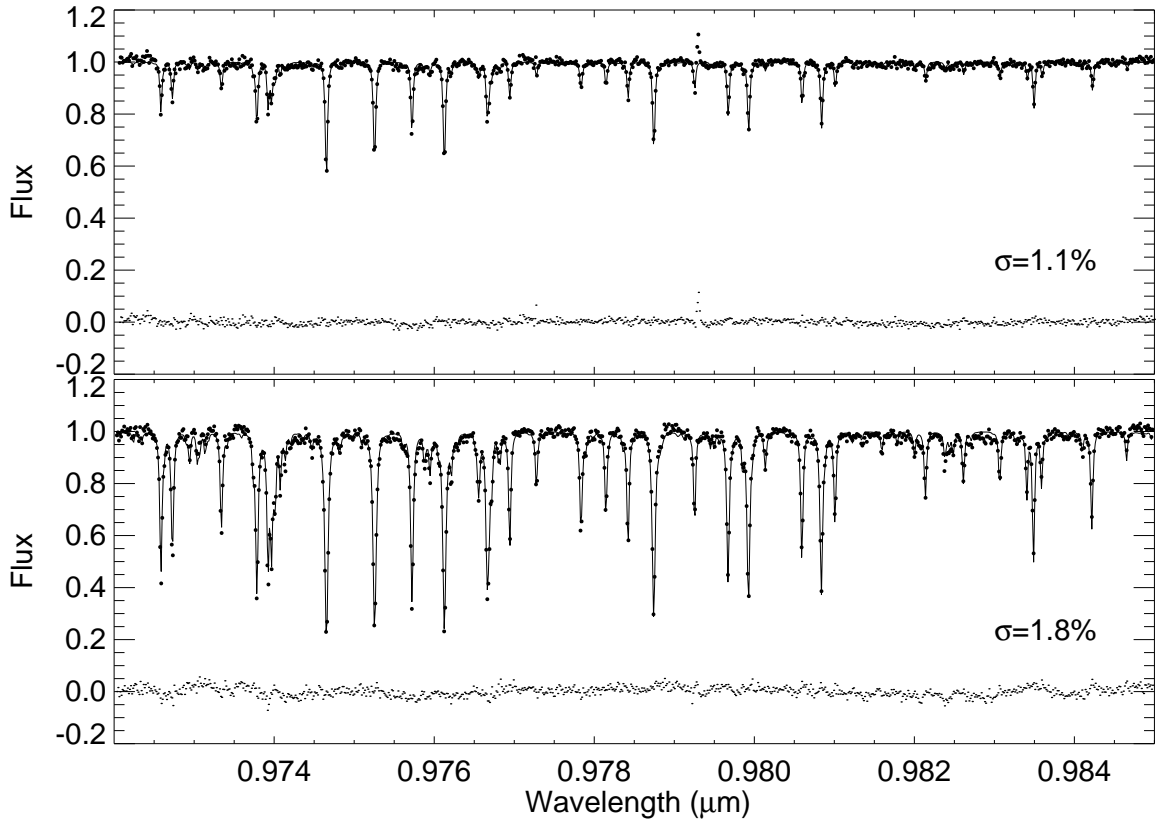}
\caption{Two examples of theoretical telluric H$_2$O templates (solid lines) fit to ARCES A star observations (small points). These spectra highlight the strong impact changes in the total H$_2$O column have on the telluric absorption. In both cases, the residuals of the fits are excellent, having scatter $<2\%$.}\label{fig1}
\end{figure}

\begin{figure}
\plotone{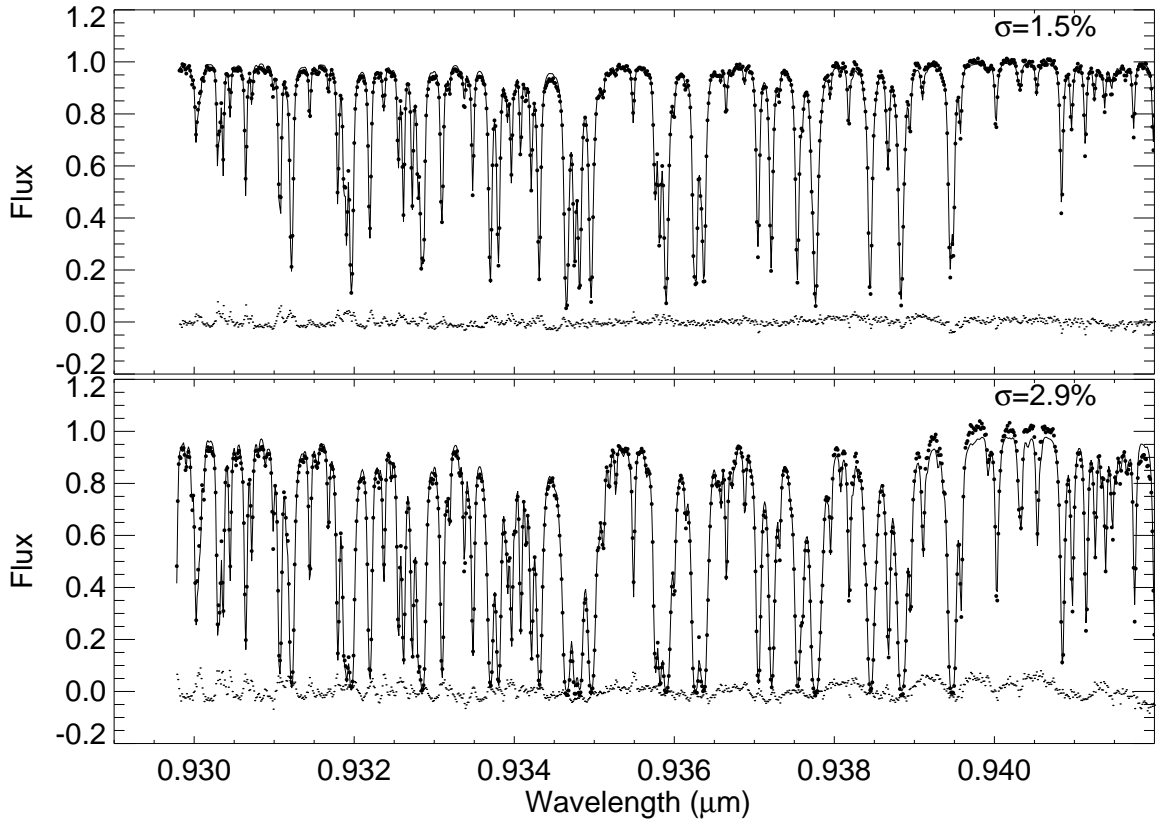}
\caption{Two examples of theoretical telluric H$_2$O templates (solid lines) fit to ARCES A star observations (small points) in a spectral region with strong H$_2$O absorption, including saturated lines. Here we see evidence for significant residuals at the cores of some lines, as well as slowly varying residuals that we attribute in part to incomplete correction for fringing in the CCD detector.}\label{fig1b}
\end{figure}

Following \citet{blake2010} we adopt a forward modeling approach to fitting our ARCES A star observations. We focus on order 3, spanning the wavelength range $970<\lambda<986$~nm, where there are numerous unsaturated H$_{2}$O lines and the A stars are intrinsically devoid of spectral features. The model, $B(i)$, begins with the atmospheric transmission spectrum based on the average Earth-GRAM model, $T_{0}$, scaled for water optical depth, $\tau$, which is a function of zenith angle (or airmass) and water content of the atmosphere. This scaled spectrum $T_{0}^{\tau}$ is convolved with an estimate of the spectrograph line spread function (LSF), interpolated onto a lower resolution pixel grid that is defined through a polynomial relation between pixel, $i$, and wavelength, $W(i)$, and finally normalized by a continuum function, $N(i)$, that is a polynomial in $i$
\begin{equation}
B(i) = \left[\left( T_{0}^{\tau}(\lambda) \otimes LSF \right)\vert_{W(i)}\right] \times N(i)  
\end{equation}
where $\otimes$ indicates convolution and $\vert$ indicates ``evaluated at''. This model has 10 free parameters: four for $N(i)$, three for $W(i)$, two for the LSF, which is a 1D Moffat function \citep{moffat1969}, and $\tau$. We fit this model to order 3 of each A star spectrum using an implementation of the AMOEBA downhill simplex method (\citealt{nelder1965}, \citealt{press1992}) to minimize $\chi^{2}$. Examples of fits are shown in Figures \ref{fig1} and \ref{fig1b}. The agreement between our theoretical telluric template and the observed spectra is excellent overall, with typical residuals of $1-2\%$ and no evidence for degradation of the fits in the centers of prominent absorption lines for regions where the line density is modest and the average line optical depth is $<1$. As shown in Figure \ref{fig1b}, in regions having numerous lines with large optical depths the overall quality of the fits deteriorates and the models exhibit significant discrepancies in the cores of some absorption features. In all cases the fit residuals do exhibit slowly varying features that we attribute to imperfect correction for CCD fringing, which has a large amplitude at these red wavelengths. Uncertainty in the HITRAN H$_{2}$O line parameters may also lead to discrepancies between our models and observations and we note that incorporating empirical line parameters, such as those from \citet{alekseeva2010a}, may help to improve the overall quality of the fits.

\section{PWV Measurements}

We found that overall the best-fit model parameters had very small scatter, demonstrating that ARCES is physically stable, with small shifts in wavelength solutions over the period of a year. However, we expect that $\tau$, which is presumably related to zenith angle and the amount of H$_{2}$O in the atmosphere, should vary widely since H$_{2}$O, unlike CH$_{4}$ \citep{blake2010}, is in-homogeneously distributed in the lower atmosphere and its concentration can change by more than $10\%$ in under an hour. As a result, our best fit values of $\tau$ spanned the ranged $0.1<\tau<0.9$. The amount of H$_2$O in the atmosphere is related to Precipitable Water Vapor (PWV), a measurable quantity that should be proportional to the optical depth to absorption by this molecule. Historically, PWV, which is critically important for mm and sub-micron observations, has been measured with balloon-borne radiosondes or differential radiometers operating in the mid-IR. In the 1990s it was demonstrated that data from a multi-band geodetic quality Global Positioning System (GPS) receiver and a high accuracy barometer can be combined to derive PWV (\citealt{bevis1992}, \citealt{bevis1994}, \citealt{rocken1995}), making these devices extremely important for studies of the time and spatial dependence of water vapor for atmospheric science applications. 

The index of refraction of the atmosphere, and therefore the speed of propagation for the microwaves used by GPS, is a strong function of pressure, temperature, and water content. Both the ionosphere and neutral atmosphere induce delays in signals from GPS satellites that are measured precisely by the receiver stations. The ionospheric delays are frequency dependent, so GPS signals at multiple frequencies (typically 1.2 and 1.6 GHz) can be used to correct for this effect. The total delay at zenith (TD) induced by the neutral atmosphere can be broken into two components: ``zenith hydrostatic delay'' (ZHD), which is due primarily to dry air, and ``zenith wet delay'' (ZWD), which is due to H$_2$O vapor. The ZHD is a slowly varying function of time and can be calculated given a barometric pressure measurement to a precision better than $0.01\%$. A measurement of the TD and an estimate of the ZHD therefore provide an estimate of the ZWD (ZWD=TD-ZHD), which is proportional to PWV with a constant of proportionality that is a function of surface temperature. While it is possible to calculate this constant of proportionality to make absolute measurements of PWV, relative measurements of PWV can be made with quoted precision better than $\sim0.2$~mm given contemporaneous surface temperature and barometric pressure measurements. 

\begin{figure}
\plotone{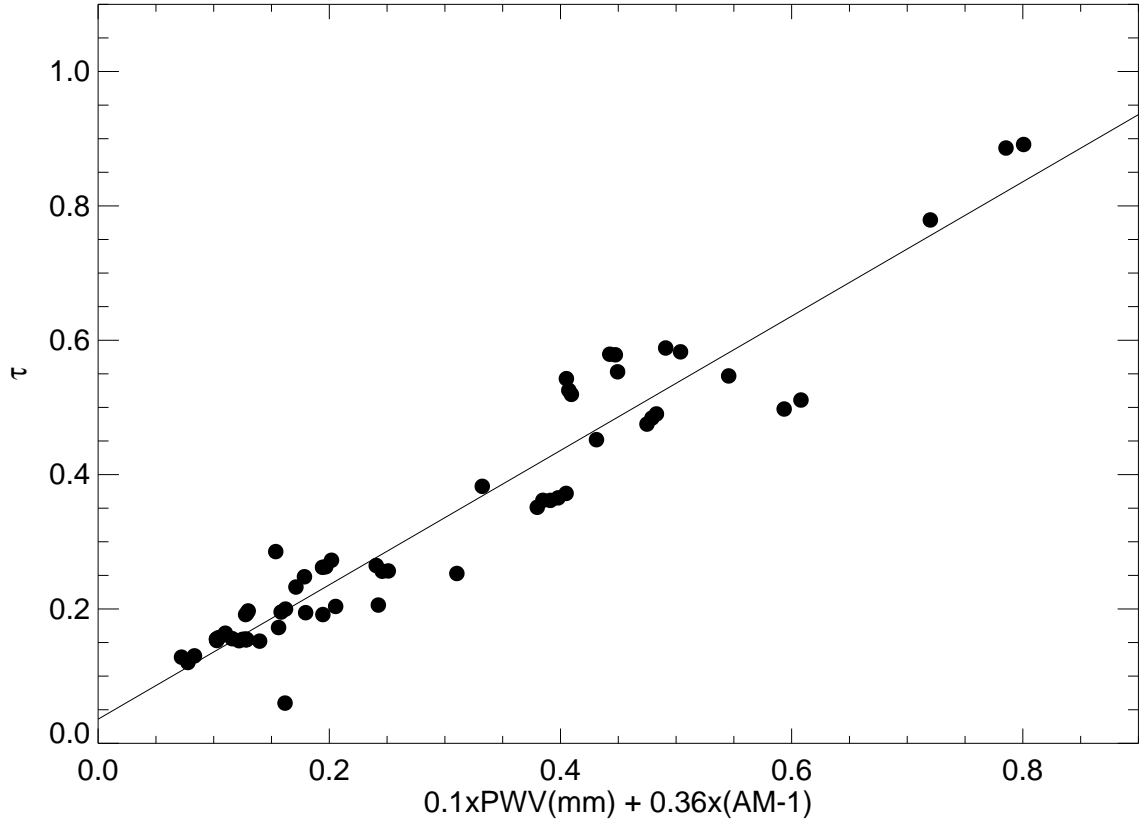}
\caption{Results of a multiple linear regression of line depth scale factor, $\tau$, against Precipitable Water Vapor and Airmass.}\label{fig2}
\end{figure}

\begin{figure}
\plotone{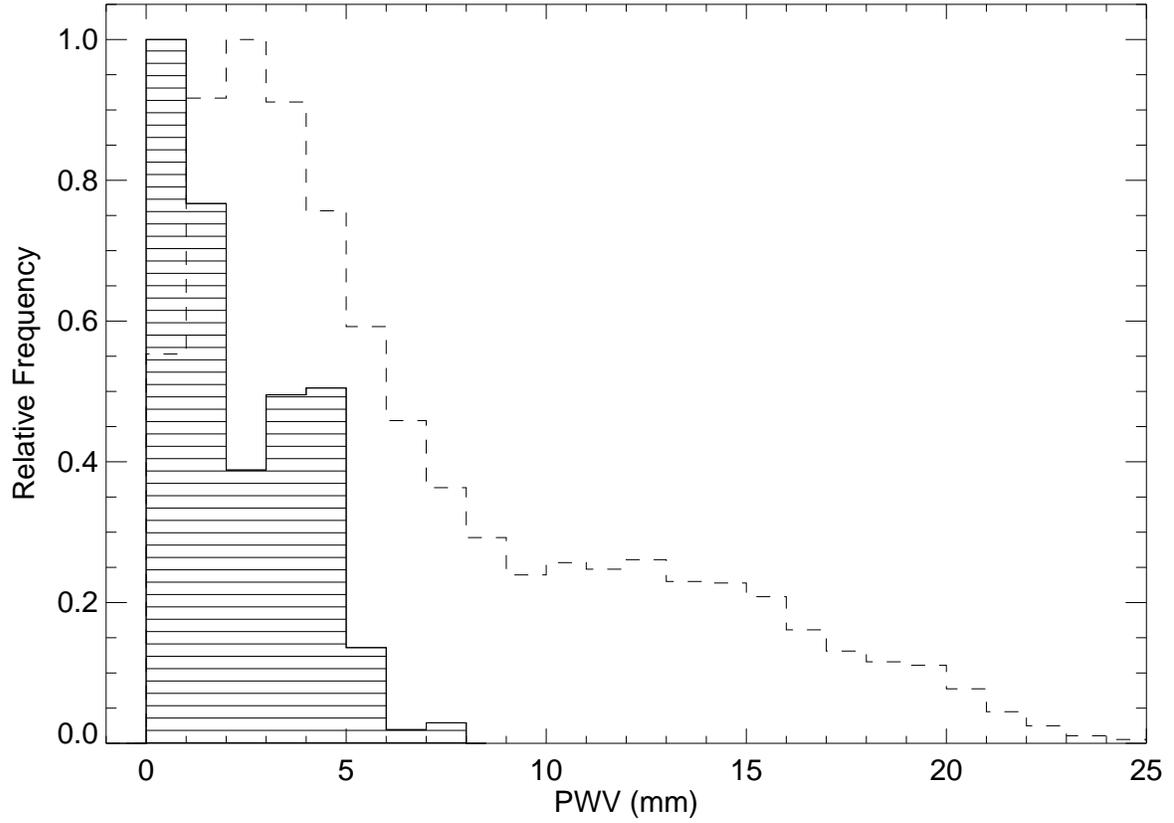}
\caption{Histogram of PWV for ARCES A star observations (shaded region) compared to the overall distribution of PWV measurements at Apache Point Observatory. Astronomical observations are typically made from APO when PWV$<10$~mm.}\label{fig3}
\end{figure}

A large number of GPS-based PWV monitors are maintained around the world. Data from several receiver networks in North America are maintained by UCAR through \textit{SuomiNet}\footnote{\url http://www.suominet.ucar.edu/}, including a station operated by UNAVCO located in close proximity to the APO 3.5~m telescope. Using the \textit{SuomiNet} web interface we downloaded PWV estimates at 30-minute intervals from the APO station for the 2009 and 2010 seasons and interpolated the PWV estimates to the times of our A star observations, excluding cases where there was not a PWV estimate within 1 hour of the time when an A star spectrum was obtained. Using multiple linear regression, we fit for our measured values of $\tau$, the scale factor specific to our telluric template, as a function of PWV and (AM-1) for PWV$<8$~mm. We found that measured PWV and AM are strongly correlated with $\tau$ and determined a best-fit linear relation
\begin{equation}
\tau = 0.036+0.098\times PWV(mm)+0.36\times(AM-1)  \pm 0.055
\end{equation}
The results of this fit are shown in Figure \ref{fig2}. We note that this result is qualitatively similar to those reported by \citet{querel2008} and \citet{thomas-osip2007} who found that PWV estimates derived from radiometer observations strongly correlated with the measured depths of H$_2$O features in optical spectra. As shown in Figure \ref{fig3}, our A star observations spanned a relatively narrow range of PWV compared to the overall distribution of PWV at APO, but we are biased toward lower PWV since periods of large PWV are likely to be unusable for astronomical observations. In practice, the coefficients of a relation like the one given in Equation 8, which is based on telluric models calculated using an ``average'' atmospheric profile for APO, will need to be calibrated once for a given observing site. Once this is done with a set of A star observations, the technique presented here should be broadly applicable to NIR observations at sites other than APO.

\section{SDSS Observations}

Our fits to the A star spectra demonstrate that given an estimate of PWV we can generate a theoretical model for atmospheric transmission in z band that has no free parameters and matches the observed telluric absorption around 980~nm to better than $2\%$ on a per-point basis. Integrated across a wide observing band, the discrepancy in total transmission between our models and the A star observations is significantly smaller, less than $0.2\%$. This opens up the possibility that GPS-based PWV estimates can be used as input to theoretical radiative transfer calculations to enable the precise estimation of molecular absorption integrated across NIR bands in real time. When combined with information about the spectra of the objects being observed, this technique could potentially lead to significant improvements in global NIR photometric precision for surveys like LSST and augment an overall calibration strategy like the ones outlined by \citet{burke2010} and \citet{jones2010}. We test this hypothesis by considering z-band photometric measurements of M stars in the SDSS photometric database. 

The SDSS \citep{york2000} made an imaging survey of the sky using a large-format CCD camera 
\citep{gunn1998} mounted on the Sloan Foundation 2.5~m telescope 
\citep{gunn2006} at APO to image the sky
in five optical bands - u, g, r, i, and z \citep{fukugita1996}.
The imaging data were reduced by a set of 
software pipelines that produced a catalog of objects with calibrated
magnitudes and positions (\citealt{lupton2002}, 
\citealt{hogg2001}, \citealt{padmanabhan2008}). SDSS Data Release 8 (DR8; \citealt{aihara2011}) includes 
photometric measurements in five bands of over 260 million stars over an area of more than 15,000 square degrees. The GPS-based PWV monitor at APO (where both the SDSS and 3.5~m telescopes are located) was installed in 2007 and began regular operations in 2008. Unfortunately, nearly all the SDSS imaging was acquired prior to this. A similar GPS-based PWV monitor operated by NOAA\footnote{\url http://www.gpsmet.noaa.gov/} is located at White Sands (WS), New Mexico, 50 km southwest of APO at an elevation of 1200 m (1580 m below APO). We interpolated the PWV estimates from WS during 2009 to the times of those from APO having PWV$<10$~mm, only considering APO measurements within 30 minutes of a WS measurement. We found that the APO and WS PWV measurements are highly correlated (Pearson's r=0.85), though with significant deviations from linearity at PWV$<1$~mm. We adopt an empirical relation between WS and APO by interpolating the average PWV data given in Table \ref{table1}. While the scatter in this relation is large, particularly at low PWV, it is sufficient for considering the relationship between the overall statistical properties of SDSS z-band photometry and PWV.

The photometric measurements in the SDSS DR8 database have undergone the calibration procedure described in \citet{padmanabhan2008}, which has been shown to achieve global photometric precision of $\sim2\%$ in z band, and $\sim1\%$ in g, r, and i bands. In z band, in addition to the components of atmospheric extinction that vary smoothly with wavelength, such as Rayleigh scattering by molecules, Mie scattering by aerosols, and ``gray'' absorption by thin clouds, line absorption by H$_2$O becomes important. This means that the extinction for a given object depends on the SED of the object, which is not the case for the other bands. \textit{\"{U}bercal} is fundamentally a differential approach, where flux is calibrated relative to a large ensemble of objects having a wide range of SEDs that, when averaged, at NIR wavelengths have a smooth SED like a solar type star. Given this, we expect systematic residuals in the SDSS z-band measurements for objects with complex NIR SEDs, such as M stars, that will correlate with PWV. We selected two samples of M stars from the SDSS DR8 database\footnote{\url http://skyserver.sdss3.org/dr8/en/tools/search/iqs.asp} to investigate this effect. 

\begin{figure}
\plotone{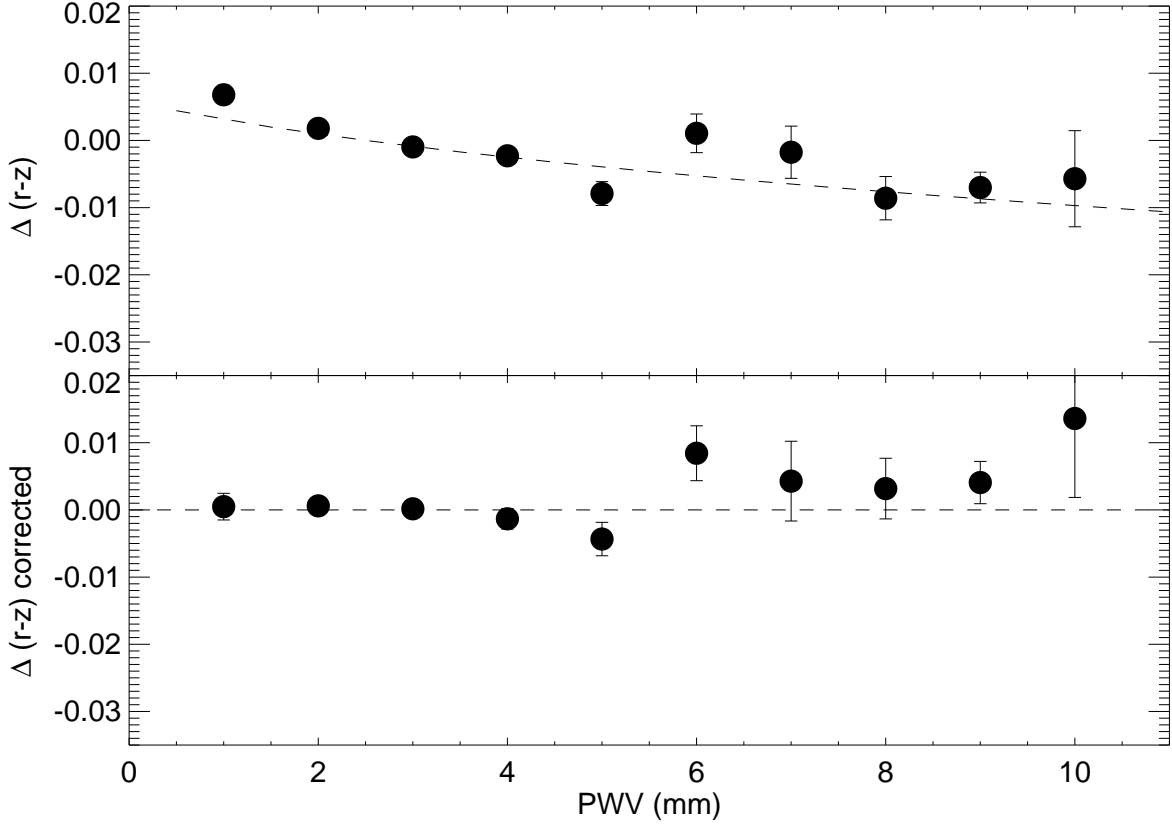}
\caption{Top panel shows mean colors of 6,177 mid-M stars binned as a function of PWV. Here, $\Delta (r-z)$ is the difference in the measured color of a star relative to the stellar locus in (r-i) vs. (r-z) space.The dashed line is the result of a theoretical prediction that takes into account telluric H$_2$O absorption scaled following Equation 8, the SDSS filter passbands, and synthetic stellar spectra from \citet{brott2005}. The theoretical (r-z) color difference is calculated between a mid-M star and a mid-G star reference, which should be representative of the average SED of all the bright stars in SDSS. Error bars are the formal statistical error on the mean of each bin and the null hypothesis of $\Delta(r-z)=0$ is ruled out at $>99\%$ confidence. Bottom panel shows results of a consistency check were an empirical $\Delta(r-z)$ vs. PWV correction is calculated from a random half of the mid-M stars and then this correction is applied to the other half of the observations. After correction, the data are consistent with $\Delta(r-z)=0$ ($P(\chi^{2})=0.73$)}\label{fig5}
\end{figure}

The first sample is based on the spectroscopically-classified catalog of M stars produced by \citet{west2011}. We selected 7,300 objects having spectral types from M7 to M4 and brighter than i=17.0 and 
retrieved the CLEAN photometric measurements of these objects from the STAR view in the DR8 context. We estimated PWV at the time of each observation based on the WS data and rejected epochs where no WS data was available. We also rejected outliers in (r-i) vs.(r-z) color space by excluding objects with (r-z) colors falling more than 0.2 mag from a linear fit to the stellar locus and calculated $\Delta (r-z)$, the color of each object relative to the stellar locus for a given (r-i). We also calculated the mean, with outlier rejection, of $\Delta(r-z)$ in each of the six camcols and found significant offsets which we subtracted off. The relation between $\Delta(r-z)$ and PWV is shown in Figure \ref{fig5}. The best fit linear relation between change in (r-z) color and PWV is $\Delta(r-z)=(-0.0019\pm0.0002)\times(PWV-3.3~\rm{mm})$~mag.

\begin{figure}
\plotone{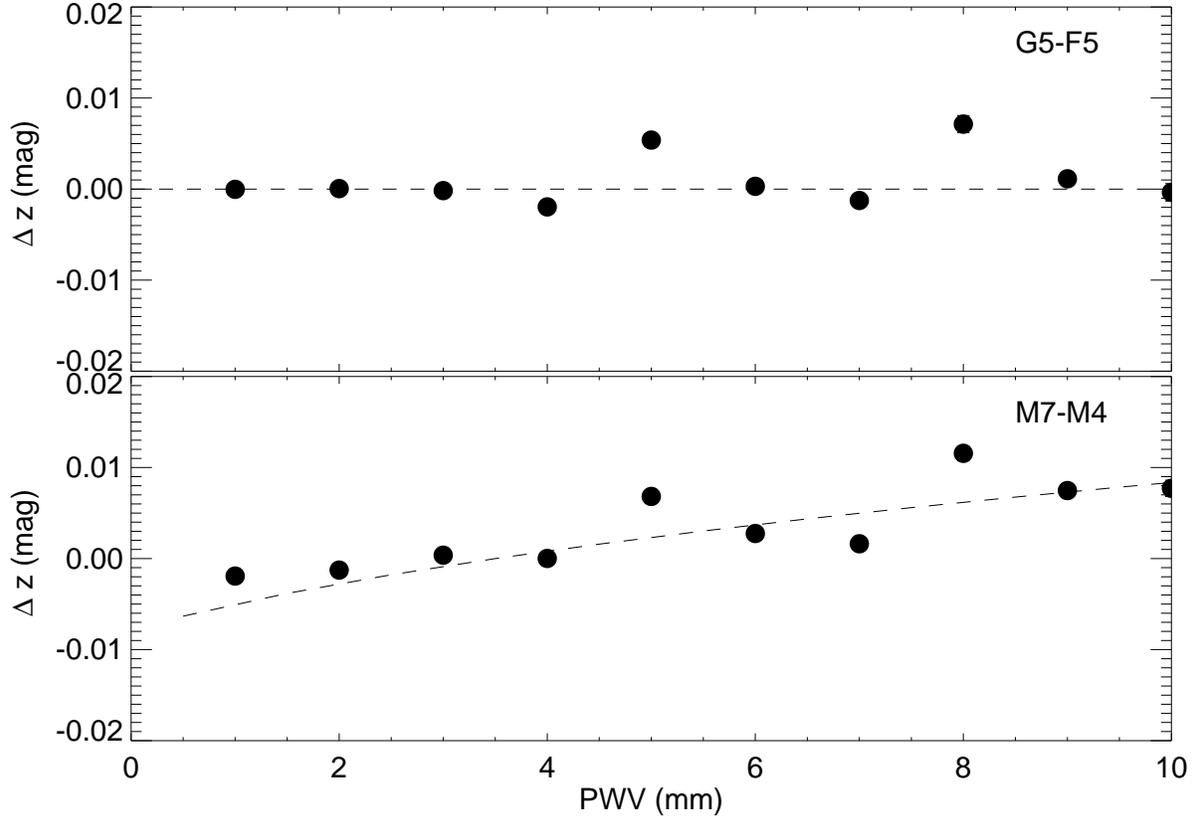}
\caption{Mean differential z-band magnitude, $\Delta z$, binned as a function of PWV for objects in Stripe 82 that were repeatedly observed by SDSS. The dashed lines are based on the same theoretical calculations shown in Figure \ref{fig5}, again assuming a mid-G star as the reference SED. Formal statistical errors on the means are smaller than the points. 14,597 observations of 983 stars are included in the top panel and 25,005 observations of 1,553 stars in the bottom panel.}\label{fig6}
\end{figure}

The second sample contains objects in Stripe 82, a 300 square degree area along DEC=0 that was repeatedly observed by SDSS \citep{ivezic2007, bramich2008, sako2008}. Based on the stellar locus defined in \citet{covey2007} we selected a random 3,000 G5-F5 stars and 2,500 M7-M4 stars from the Stripe 82 standard star catalog of \citet{ivezic2007}, restricting our samples to objects brighter than z=17.0. While parts of Stripe 82 have been observed up to 80 times, only a portion of those visits are included in DR8 \citep{aihara2011}. While all the visits are included in DR7, these data were not globally calibrated using \textit{\"{U}bercal}.
We selected photometric observations of our Stripe 82 targets from the DR8 PHOTOOBJ view requiring all the photometric flags required for CLEAN status except for PRIMARY. We culled observations at epochs where no WS PWV data was available, as well as observations with colors offset more than 0.1 mag from the (g-i) stellar locus defined in \citet{covey2007}. In cases where observations of the same object spanned multiple camcols, we chose a single camcol for that object and rejected observations from other camcols. We also excluded objects with fewer than five remaining observations at PWV$<4$~mm. The final samples contain 14,597 observations of 983 G5-F5 stars and 25,005 observations of 1,553 M7-M4 stars. For each object in each sample we calculate $\left<z\right>_{4}$, the average of the z-band magnitudes only for PWV$<4$~mm, and then $\Delta z=z-\left<z\right>_{4}$. The relation between $\Delta z$ and PWV for both samples is shown in Figure \ref{fig6}. As expected, the effect of differential extinction due to H$_2$O is small for G and F stars, which have SEDs similar to the SED of the average point source used in the photometric calibration, but the effect is large for M stars, with $\Delta z_{H_{2}O}=(0.0012\pm0.00008) \times (PWV-2.6~\rm{mm})$~mag. The slope of this relation is consistent with that of the $\Delta(r-z)$-PWV relation, indicating that our results obtained using two different sets of observations are very similar. Considering that the distribution of PWV for the Stripe 82 observations is roughly uniform from 1 mm to 10 mm, this is potentially an important component of the global calibration of SDSS M star observations. However, taking the standard deviation of the empirical corrections for all the SDSS measurements indicates that this is a small effect overall: $\left<\Delta z_{H_{2}O}\right>=0.003$~mag. For the M stars we recalculated $\Delta z$ after applying this empirical PWV correction. Using the F-statistic \citep{press1992} we compared the variances of $\Delta z$ before and after the PWV correction and found no statistically significant reduction in the variance of $\Delta z$ ($P(\sigma^{2}_{1}= \sigma^{2}_{2})=0.44$). 

\begin{figure}
\plotone{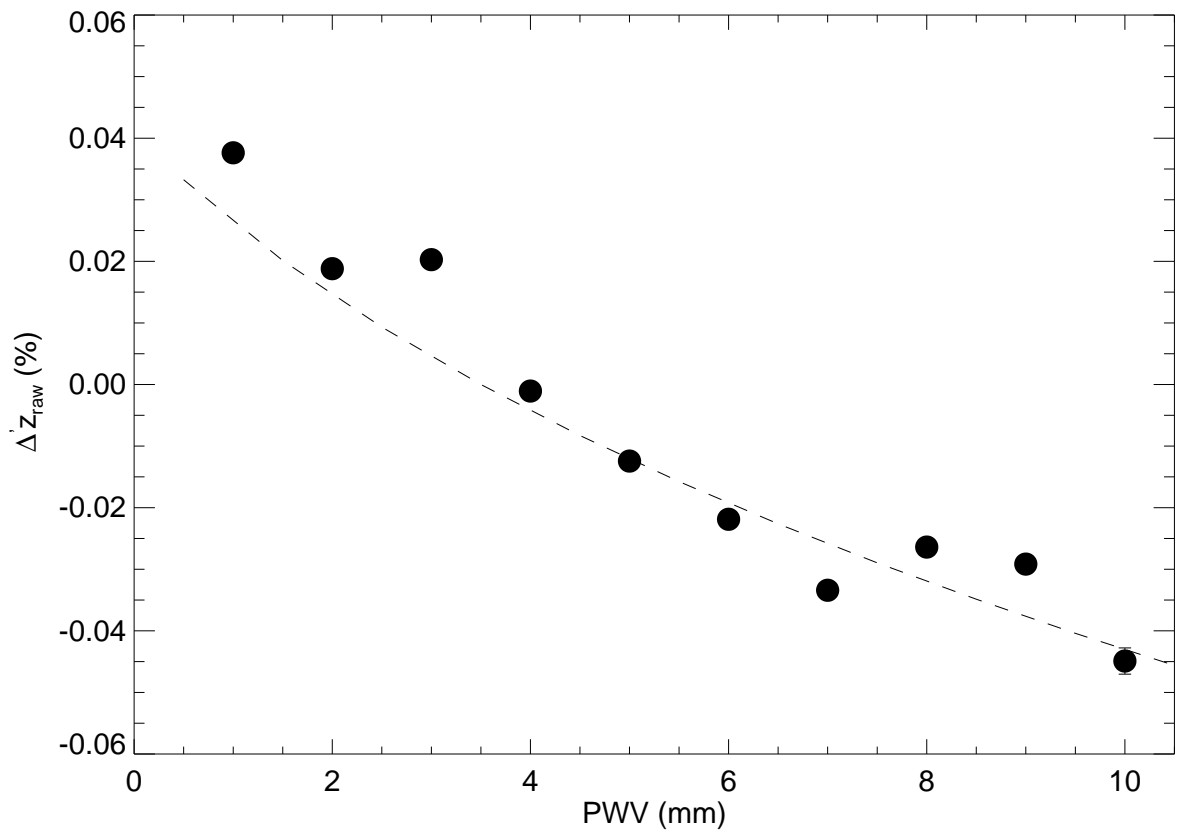}
\caption{Mean values of 14,597 uncalibrated z-band flux measurements for the Stripe 82 G5-F5 star sample binned as a function of PWV. As discussed in Section 6, $\Delta^{'}z_{raw}$ is an estimate of the change in raw z-band flux after correction for extinction as measured simultaneously in r band. The dashed line is based on the same theoretical calculations shown in Figure \ref{fig5}. The changes in total absorption in z-band with PWV are significant, up to several percent, but these are largely removed via relative calibration techniques, with residuals that depend on SED. Statistical errors on the means are smaller than the points. }\label{fig7}
\end{figure}

Using the Stripe 82 sample we can also investigate changes in overall z-band transmission with PWV by considering uncalibrated flux measurements. Using the PSFFLUX and NMGYPERCOUNT entries in the PHTOTOOBJ view we can convert the calibrated fluxes retrieved from the DR8 database back into uncalibrated, raw fluxes. Using the same procedure as for the calibrated Stripe 82 data, for each object in the G5-F5 sample we calculate $\Delta z_{raw}=z_{raw}/\left<z_{raw}\right>_{4}$ and $\Delta r_{raw}=r_{raw}/\left<r_{raw}\right>_{4}$. As a result of cloud cover and variations in the optical depth of molecular scattering, there are variations at the level of $\sim10\%$ in the uncalibrated r and z fluxes, but $\Delta r_{raw}$ and $\Delta z_{raw}$ are highly correlated. We fit a linear relation, $F(\Delta r_{raw})$, for $\Delta z_{raw}$ as a function of $\Delta r_{raw}$ and then calculate $\Delta^{'} z_{raw}=\Delta z_{raw}-F(\Delta z_{raw})$. Since the SDSS r band is free from strong H$_2$O line absorption and probes instead the transmission variations due to scattering, we use $\Delta^{'} z_{raw}$ as a proxy for the molecular component of absorption in z band. The change in raw z flux as a function of PWV is shown in Figure \ref{fig7}. The best fit linear relation between molecular absorption and PWV is $\Delta^{'} z_{raw}(\%)=(-0.0096\pm0.0001)\times(PWV-4.3~\rm{mm})$.

We compared these empirical relations between z raw flux, z magnitude, and the (r-z) colors of M stars to theoretical estimates of the impact of changing PWV on SDSS z-band transmission. We numerically integrated the synthetic stellar spectra described by \citet{brott2005} across the SDSS passbands\footnote{\url http://www.sdss3.org/instruments/camera.php} including our H$_2$O absorption template for APO scaled following our derived $\tau(PWV)$ relation given in Equation 8 and assuming AM=1.2. We focused on stellar templates representative of a mid-M dwarf (T$_{\rm{eff}}$=3000 K, $\log(g)$=5.0) and a mid-G dwarf (T$_{\rm{eff}}$=6000 K, $\log(g)$=4.5). For values of PWV in the range 0.5 to 12 we calculated the expected z-band and r-band fluxes for each template as a function of PWV, normalizing to the expected flux at PWV=3 mm. While the expected z-band fluxes change by more than $8\%$ when PWV increases from 1 mm to 10 mm, to first approximation these variations are removed by the SDSS photometric calibration. If the appropriately weighted average of the SEDs of all SDSS stars is well-approximated by a G-star spectrum, then the differential fluxes we calculate in r and z bands should match the variations we measure in SDSS z-band photometry as a function of PWV. These theoretical expectations are shown as dashed lines in Figures \ref{fig5} and \ref{fig6} and are in very good agreement with our measurements. We experimented with including different Aerosol and Rayleigh scattering models in these calculations, but as functions that vary smoothly with wavelength they had a negligible effect on the predicted differential M dwarf colors, fluxes, and differential magnitudes.

\section{Discussion}

\begin{figure}
\plotone{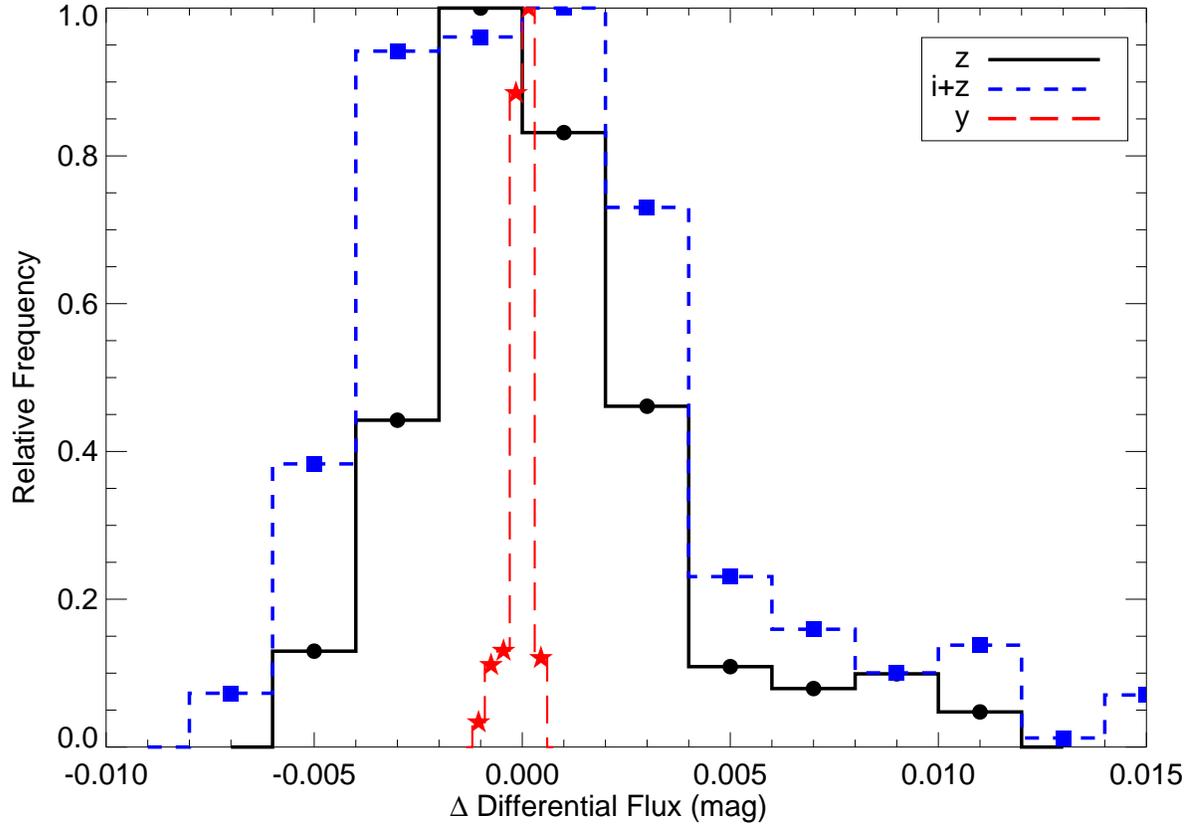}
\caption{Histograms of the expected variation in differential magnitude for a mid-M target and mid-G comparison star in different NIR bands given the PWV values at the times of all of the SDSS observations in our M star sample. The standard deviations of these distributions are: $\sigma_{\Delta z}=0.003$, $\sigma_{\Delta i+z}=0.004$, and $\sigma_{\Delta y}=0.0003$~mag.}\label{fig9}
\end{figure}

Empirical corrections for extinction due to H$_2$O absorption may be important for a wide range of spectroscopic and photometric astronomical observations at deep red and infrared wavelengths. In many cases, contemporaneous PWV measurements derived from GPS may provide an important external parameter that can be used to correct for these effects. Precise differential measurements of stellar brightness can be used to find rocky planets transiting low-mass stars \citep{blake2008a, nutzman2008, charbonneau2009} as well as measure the properties of the atmospheres of these planets in transmission \citep{bean2011, crossfield2011,croll2011}. For these observations, precision at the level of 1 part in $10^{3}$ or better is the objective. At this level, second-order extinction effects stemming from variations in molecular absorption may be a dominant source of noise, particularly for wide NIR photometric bands. Given a mid-M target and an ensemble of G-type reference stars, the theoretical (dashed) curve in the bottom panel of Figure \ref{fig6} indicates the dependence of this effect on PWV.
We simulated the impact of PWV variations on a long-term M dwarf survey by calculating the differential flux for a mid-M star and a mid-G comparison star given the overall distribution of PWV at APO corresponding to the times of 6,200 SDSS M star observations. The results of this calculation, shown in Figure \ref{fig9} indicate that for the SDSS z band and a wide i+z band \citep{nutzman2008}, the expected size of this effect is 0.003 to 0.004 mag, comparable to the signal from a super-Earth transiting a mid-M star. Large changes in PWV can occur on rapid timescales, comparable to the duration of a transit event, further complicating matters. A multiple linear regression of the differential flux of an M star target relative to the ensemble reference against PWV and AM may reveal significant correlation. The coefficients of this correlation are uniquely determined by the SEDs of the target and the reference stars and could be used as an empirical relation to improve photometric measurements by correcting for the impact of variations in H$_2$O absorption (see \citealt{burke2011}).

Future wide-area photometric surveys like the Large Synoptic Survey Telescope (LSST; \citealt{ivezic2008}) will use sophisticated strategies for calibrating the absolute transmission of Earth's atmosphere in real time. \citet{burke2010} demonstrate that spectrophotometric observations of a grid of standard stars can be used to calibrate band-integrated atmospheric transmission to much better than $1\%$ by modeling both scattering and absorption. While absorption by H$_2$O is just one component of the complex and variable atmospheric transmission function, our results indicate that GPS-based PWV measurements may augment spectrophotometric observations of standard stars, narrow-band photometric observations at key wavelengths \citep{alekseeva2010b}, or an overall internal calibration procedure \citep{jones2010} and help improve photometric precision for objects with strong spectral features in the NIR. Another approach to mitigating the effect of variable H$_2$O absorption is to design filters that avoid prominent absorption bands, such as those between 900 and 1000 nm, though at the cost of the total number of stellar photons detected. Some possible designs for the LSST y band filter incorporate a blue edge that avoids H$_2$O absorption \citep{high2010}. We simulated the impact of variable PWV on differential measurements of a mid-M star and a mid-G comparison star using the y$_3$ filter\footnote{\url http://ssg.astro.washington.edu/elsst/magsfilters.shtml?filterinfo} shown in Figure \ref{fig8}. The results of this simulation, shown in Figure \ref{fig9}, indicate that careful selection of NIR filters can help suppress the impact of variable PWV on M star photometry to below 1 mmag.

Careful modeling of telluric lines is critically important for deriving precise stellar radial velocities from high-resolution NIR spectra. \citet{wang2011} showed that to approach the theoretical photon limit in Doppler measurements of low-mass stars, telluric lines need to be modeled with residual depths $<1\%$. While several methods exist for empirical estimation of telluric models using observations of A stars (e.g. \citealt{vacca2003}), our results indicate that theoretical  telluric templates combined with GPS-based PWV data can also be used for this purpose. Telluric templates that match observed line depths to better than $1\%$ may be an important component of the analysis of data from a new generation of high-resolution NIR Doppler instruments (e.g. \citealt{mahadevan2010}).

\section{Conclusion}

We calculate theoretical telluric transmission templates in the deep red appropriate for Apache Point Observatory (APO) and demonstrate that these models, which contain almost exclusively absorption by H$_2$O, are excellent fits to high-resolution observations of A stars. These results are qualitatively similar to those from \citet{thomas-osip2007}, \citet{querel2008} and \citet{seifahrt2010} who fit theoretical models to observations of a wide range of molecular absorption features in the optical and IR. We compare the scaling parameter for the depths of the H$_2$O absorption features around 980~nm, $\tau$, to contemporaneous estimates of Precipitable Water Vapor (PWV) derived from a Global Positioning System (GPS) receiver located at APO and find that they are strongly correlated. A multiple linear regression of PWV and airmass (AM) against $\tau$ indicates that the relative optical depth of H$_2$O can be predicted to better than $10\%$, leading to band-integrated estimates of total water absorption that are better than $0.2\%$. For regions where absorption features have modest optical depth and are well-separated, the agreement between the models and telluric absorption features in the A star spectra is better than $2\%$ even in the cores of lines. For regions with dense forests of lines, leading to saturation, residuals exist at the level of $5-10\%$ within the cores of some lines. 

The real-time characterization of atmospheric molecular absorption is important for the calibration of broadband NIR photometry. Even using sophisticated differential global calibration techniques, changes in PWV can bias measurements of the fluxes of objects having strong spectral features in the NIR given that the average SED of the reference ensemble of stellar objects is likely to be much smoother. We show that globally calibrated SDSS measurements of the z-band fluxes and r-z colors of mid-M stars are correlated with PWV. This is not the case for G and F stars, since their NIR SEDs are similar to that of the ``average'' star included in the \textit{\"{U}bercal} photometric calibration. We compare the observed relation between PWV and the M dwarf photometric measurements to theoretical predictions based on transmission of the SDSS filters, synthetic stellar templates, and our telluric models and find that they are in excellent agreement. We measure the size of the impact of changing PWV on SDSS z-band measurements of mid-M stars to be $\Delta z_{H_{2}O}=(0.0012\pm0.00008)\times (PWV-2.6~\rm{mm})$~mag. We applied this empirical correction to repeat observations from Stripe 82, but found no statistically significant reduction in the variance of 25,000 relative z-band measurements of mid-M stars. Given the range of PWV during the Stripe 82 observations, the expected size of the correction is $\left<\Delta z_{H_{2}O}\right>=0.003$~mag, significantly smaller than the photon errors for the majority of stars in our sample.

GPS-based estimates of H$_2$O absorption may be an important component of the calibration strategy for future surveys like LSST and, when combined with atmospheric monitoring \citep{burke2010} or internal calibration \citep{jones2010} and a strategy for precisely measuring the spectral response of the instrument (\citealt{stubbs2006}, \citealt{stubbs2010}), may help to improve differential photometric measurements for objects with complex NIR SEDs. In the near term, modeling H$_2$O absorption through PWV measurements may help to improve the precision of differential NIR photometry of bright M stars, enabling transiting planet surveys to improve their sensitivity to small planets. At the same time, precise models of telluric absorption may enable Doppler measurements in new wavelength regimes, enhancing ongoing efforts to find Earth-like planets orbiting low-mass stars.

\acknowledgments
We thank an anonymous referee for thoughtful comments that helped to focus and improve this work. We would also like to thank Robert Lupton, Chris Stubbs, and Ji Wang for helpful conversations that contributed to this work. We would like to thank the members of the Natural Environments Branch at Marshall Space Flight Center for developing and supporting the Earth-GRAM model and for making it available to us. CHB thanks the NSF for support through an Astronomy and Astrophysics Postdoctoral Fellowship (AST-0901918). CHB would also like to thank George Rybicki for teaching an excellent class about radiative processes. This work is based in part on observations obtained with the Apache Point Observatory 3.5-meter telescope, which is owned and operated by the Astrophysical Research Consortium. We thank UNAVCO for installing and operating the GPS station at Apache Point and the \textit{SuomiNet} project at UCAR for producing  real-time PWV estimates for this station. Funding for SDSS-III has been provided by the Alfred P. Sloan Foundation, the Participating Institutions, the National Science Foundation, and the U.S. Department of Energy Office of Science. The SDSS-III web site is \url{http://www.sdss3.org/}.
SDSS-III is managed by the Astrophysical Research Consortium for the Participating Institutions of the SDSS-III Collaboration including the University of Arizona, the Brazilian Participation Group, Brookhaven National Laboratory, University of Cambridge, University of Florida, the French Participation Group, the German Participation Group, the Instituto de Astrofisica de Canarias, the Michigan State/Notre Dame/JINA Participation Group, Johns Hopkins University, Lawrence Berkeley National Laboratory, Max Planck Institute for Astrophysics, New Mexico State University, New York University, Ohio State University, Pennsylvania State University, University of Portsmouth, Princeton University, the Spanish Participation Group, University of Tokyo, University of Utah, Vanderbilt University, University of Virginia, University of Washington, and Yale University.

{\it Facilities:} \facility{Apache Point Observatory}.

\bibliography{mybib}{}

\clearpage

\clearpage

\begin{deluxetable}{rrr}
\tablecaption{Empirical relation between PWV measured at White Sands and PWV measured at APO.\label{table1}}
\tablewidth{0pt}
\tablehead{
\colhead{PWV WS} & \colhead{PWV APO} & \colhead{$\sigma_{APO}$}\\
\colhead{mm} & \colhead{mm} & \colhead{mm}
}
\startdata
1.83 & 0.66  & 0.40 \\
2.44 & 0.80 & 0.50 \\
3.26 & 1.29 & 0.71  \\
4.34 & 1.98 & 0.84\\
5.79 & 2.77 & 1.01 \\
7.72 & 3.80 & 1.20 \\
10.30 & 5.02 & 1.52 \\
13.72 & 6.60 & 1.66 \\
18.30 & 8.79 & 2.10 \\
24.41 & 12.38 & 2.48 \\

\enddata
\end{deluxetable}

\end{document}